\documentclass[12pt, a4paper, twoside]{article}
\usepackage[utf8]{inputenc}
\usepackage[english]{babel}
\usepackage[a4paper, left=2cm, right=2cm, top=2cm, bottom=2cm]{geometry}
\usepackage{amsmath}
\usepackage{amssymb}
\usepackage{graphicx}
\usepackage{float}
\usepackage{wrapfig}
\usepackage{pdfpages}
\usepackage[skip=2pt,font=scriptsize]{caption}
\usepackage{cite}	
\usepackage{appendix}
\usepackage{pdfpages}
\usepackage{url}
\usepackage{slashed}
\usepackage{authblk}
\usepackage{textcomp}
\usepackage[T1]{fontenc}

\title{KWISP: an ultra-sensitive force sensor for the Dark Energy sector}

\date{\today}

\begin{document}

\author[1,3]{M.~Karuza}
\affil[1]{\footnotesize Phys. Dept. and CMNST, University of Rijeka, R. Matejcic 2, Rijeka, Croatia}
\author[2,3]{G.~Cantatore \thanks{giovanni.cantatore@ts.infn.it}}
\affil[2]{\footnotesize Università di Trieste, Via Valerio 2, 34127 Trieste, Italy} 
\affil[3]{\footnotesize INFN Sez. di Trieste, Via Valerio 2, 34127 Trieste, Italy} 
\author[4]{A.~Gardikiotis}
\affil[4]{\footnotesize University of Patras, GR 26504 Patras, Greece}
\author[5]{D.H.H.~Hoffmann}
\affil[5]{\footnotesize Institut für Kernphysik, TU-Darmstadt, Schlossgartenstr. 9, D-64289 Darmstadt, Germany}
\author[6]{Y.K.~Semertzidis}
\affil[6]{\footnotesize Department of Physics, KAIST, Daejeon 305-701, Republic of Korea}
\author[4,7]{K.~Zioutas}
\affil[7]{\footnotesize European Organization for Nuclear Reseach (CERN), Gèneve, Switzerland}

\maketitle

\abstract{An ultra-sensitive opto-mechanical force sensor has been built and tested in the optics laboratory at INFN Trieste. Its application to experiments in the Dark Energy sector, such as those for  Chameleon-type WISPs, is particularly attractive, as it enables a search for their direct coupling to matter. We present here the main characteristics and the absolute force calibration of the KWISP (Kinetic WISP detection) sensor. It is based on a thin Si$_{3}$N$_{4}$ micro-membrane placed inside a Fabry-Perot optical cavity. By monitoring the cavity characteristic frequencies it is possible to detect the tiny membrane displacements caused by an applied force. Far from the mechanical resonant frequency of the membrane, the measured force sensitivity is $5.0 \cdot 10 ^{-14}~\mbox{N}/ \sqrt{\mbox{Hz}}$, corresponding to a displacement sensitivity of $2.5 \cdot 10 ^{-15}~\mbox{m}/ \sqrt{\mbox{Hz}}$, while near resonance the sensitivity is $1.5 \cdot 10 ^{-14}~\mbox{N}/ \sqrt{\mbox{Hz}}$, reaching the estimated thermal limit, or, in terms of displacement, $7.5 \cdot 10 ^{-16}~\mbox{m}/ \sqrt{\mbox{Hz}}$.} These displacement sensitivities are comparable to those that can be achieved by large interferometric gravitational wave detectors.

\section{Introduction}

\begin{figure}
	\centering
	\includegraphics[keepaspectratio=true, width= 15 cm]{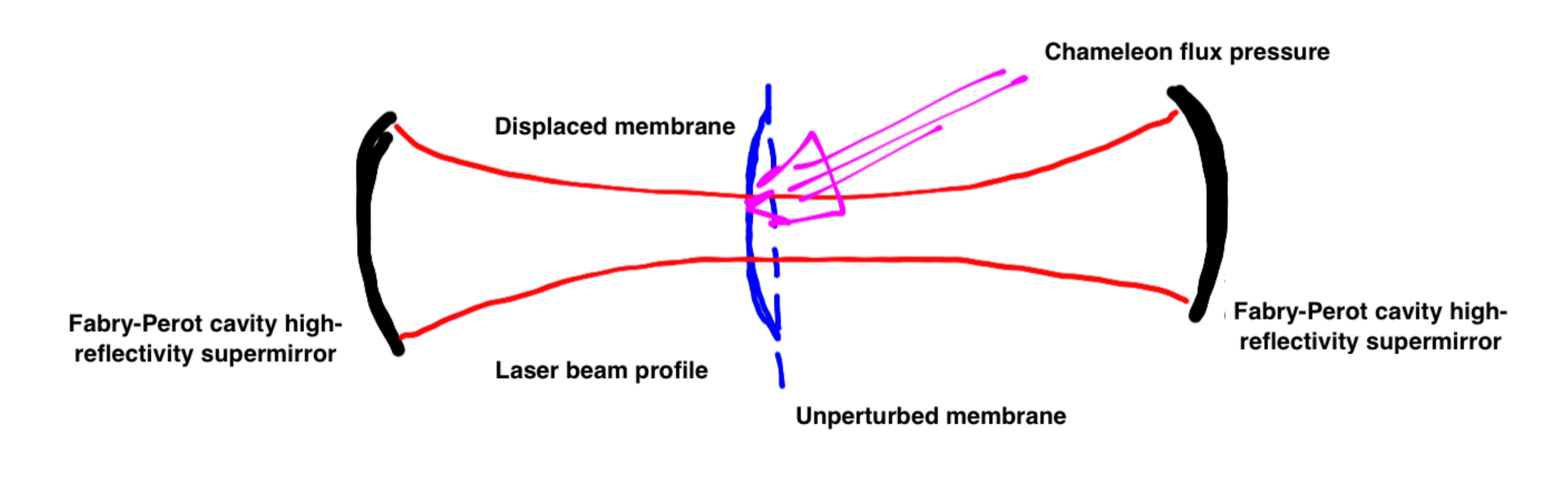}
	\caption{Sketch of the KWISP sensor working principle. The membrane flexes under the action of an external force and perturbs the resonance configuration of the intra-cavity electric field (here represented pictorially with an arbitrary beam profile). This in turn causes a shift in the cavity resonant frequencies, which can then be sensed by the feedback keeping laser and cavity in lock (see also text)}
	\label{fig:figure1}
\end{figure}

We have developed an ultra-sensitive opto-mechanical force sensor that can be applied, among other things, to searches of WISP-type particles (Weakly Interacting Slim Particles). In particular, we intend to shortly use this sensor, called KWISP for \textquotedbl Kinetic WISP detection\textquotedbl, to detect the hypothetical flux of Chameleons produced in the sun by exploiting their local density-dependent direct coupling to matter. A flux of solar Chameleons will exert the equivalent of a radiation pressure when impinging at a grazing incidence angle on a solid surface \cite{Baker2012,Baum2014}.
The KWISP sensor consists of a thin and rigid dielectric membrane suspended inside a resonant optical Fabry-Perot cavity. The collective force exerted by solar Chameleons bouncing off the membrane surface excites its vibrational states and causes a displacement from its equilibrium position. If a laser beam is frequency-locked to the cavity by means of an active electro-optical feedback system \cite{Cantatore1999}, a membrane displacement from the initial position will cause cavity mode frequencies to experience a shift \cite{Karuza2012,Karuza2013}, which is then sensed in the feedback correction signal. The sensor thus transduces displacement (force) into an electrical signal with a gain proportional to the finesse of the Fabry-Perot cavity. Figure \ref{fig:figure1} shows a pictorial representation of the KWISP sensor working principle.
The displacement sensitivity can be enhanced by exploiting the fact that the membrane is a mechanical resonator with a large figure of merit (Q factor): if an external force acts on it at the resonant frequency, resulting displacements are amplified by Q .

After a description of the sensor itself, we will present measurements done in the INFN Trieste optics laboratory to characterise it and determine its sensitivity with the direct application of an external force generated by the radiation pressure of an auxiliary laser beam. In the conclusions we will briefly expound on the perspective applications to solar Chameleon searches and to the study of short distance interactions.

\section{The KWISP force sensor}

\begin{figure}
	\centering
	\includegraphics[keepaspectratio=true,height=10 cm, angle=270]{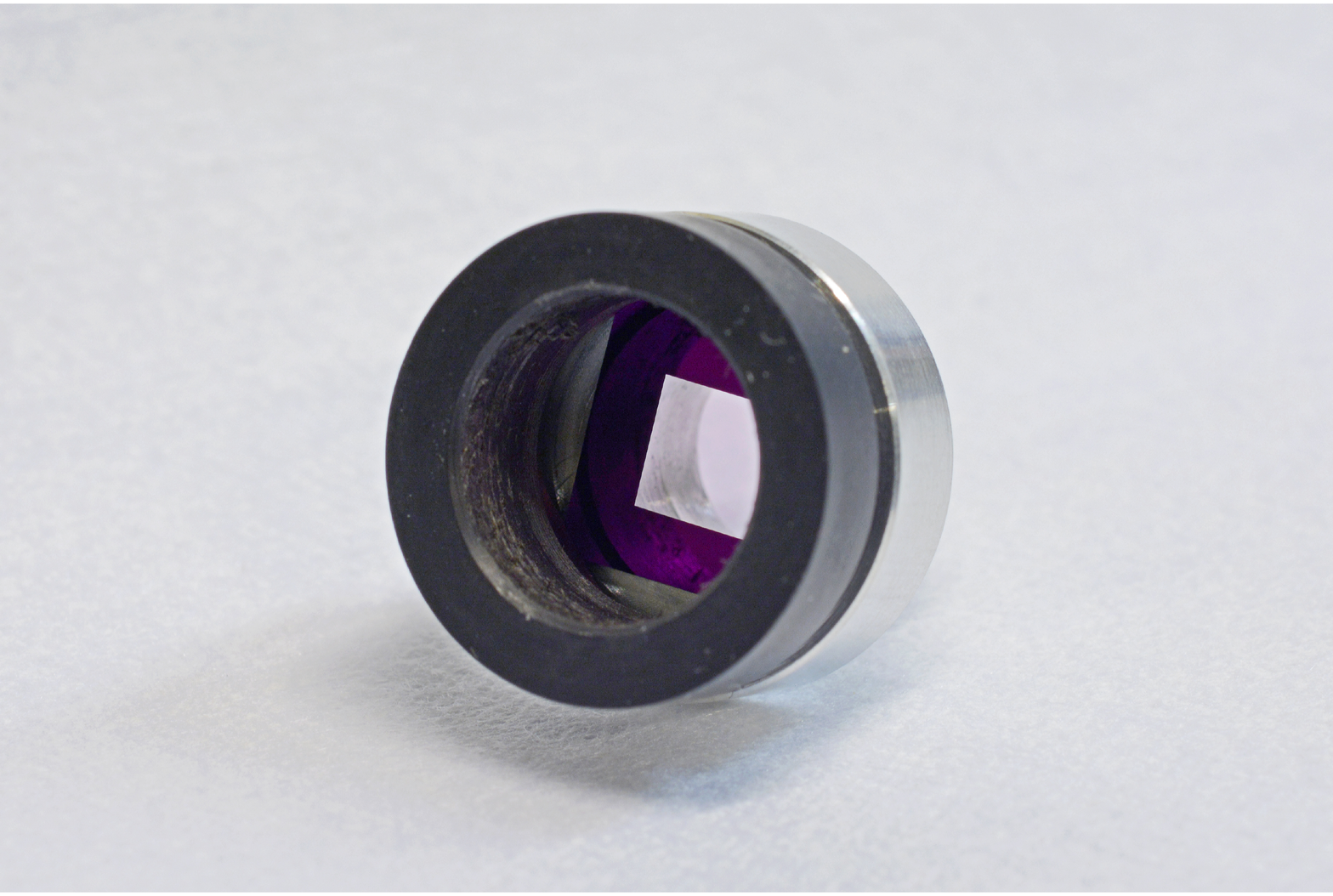}
	\caption{Photograph of the membrane holder. The Si$_{3}$N$_{4}$ membrane itself is visible as a square-shaped window inside the holder. Membrane dimensions: (5 mm)x(5 mm)x(100 nm).}
	\label{fig:figure2}
\end{figure}

A KWISP force sensor is presently installed in the optics laboratory at INFN Trieste. The main element of the sensor is a vacuum chamber containing an 85 mm long Fabry-Perot cavity made with two 1-inch diameter, 100 cm curvature radius, high-reflectivity, multilayer dielectric mirrors (made by ATFilms, Boulder, Co., USA). Each mirror is mounted on a two-axis, piezo-actuated, tilting mount (Agilis series by Newport, USA), which is in turn fixed to a common base. 
A  Si$_{3}$N$_{4}$, 5x5 mm$^{2}$, 100 nm thick membrane (made by Norcada Inc., Canada) is inserted in a holder mounted on a 5-axis movement stage, allowing movement of the membrane along 3 linear axes, one of which parallel to the cavity axis, and tilting of it around two additional axes. Using this mechanical assembly, the membrane is initially placed approximately midway between the two cavity mirrors (\emph{membrane-in-the-middle} configuration).
Figure \ref{fig:figure2} shows a photograph of the membrane holder with the (5 mm)x(5 mm) membrane inside.

\begin{figure}
	\centering
	\includegraphics[keepaspectratio=true,height=12 cm]{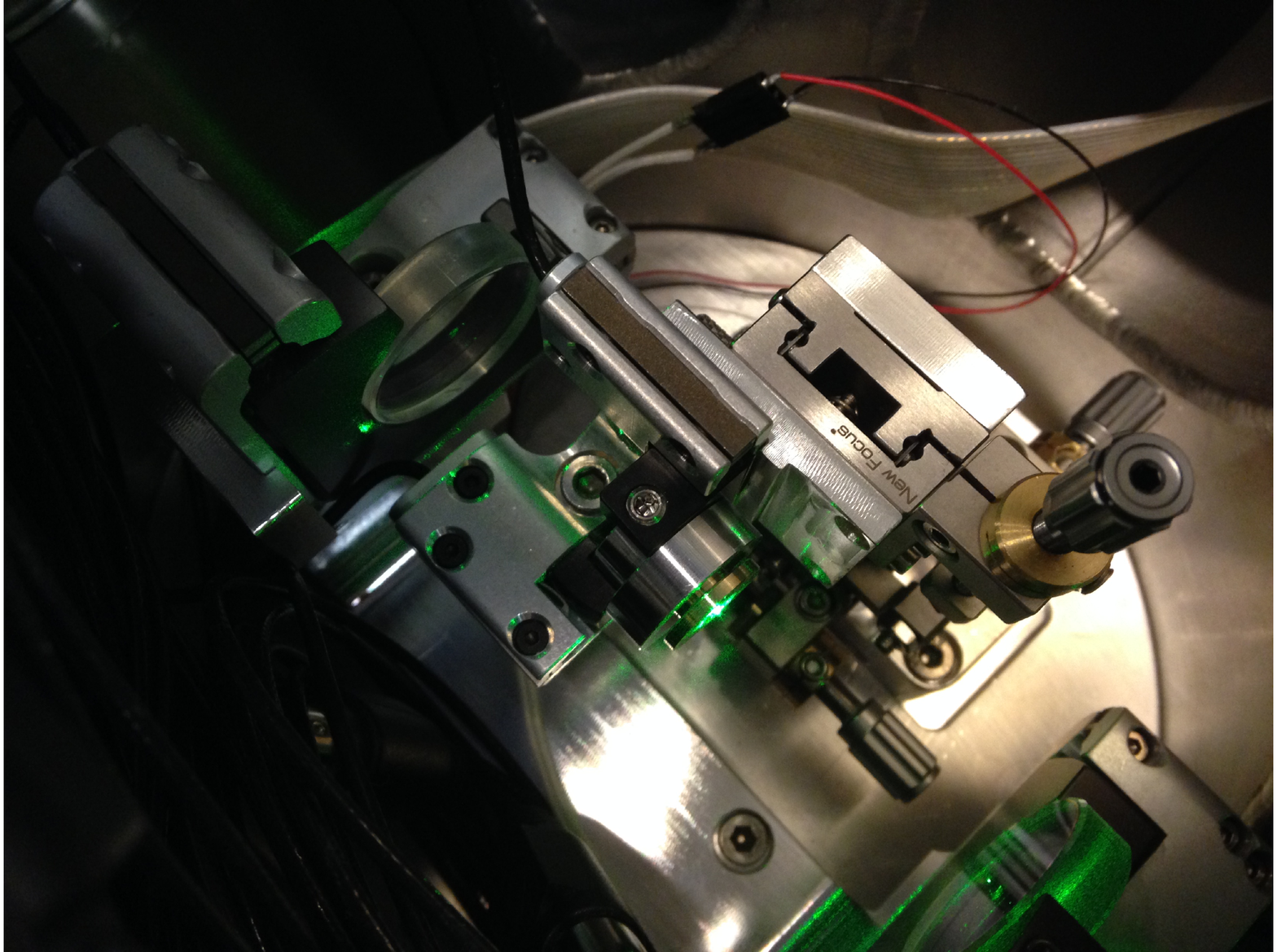}
	\caption{Photograph of the interior of the KWISP sensor vacuum chamber. The membrane holder fixed on its 5-axis mount is visible at center. Cavity mirrors are visible at upper left and lower right. An auxiliary green light beam is used to highlight the main components (see also text).}
	\label{fig:figure3}
\end{figure}

The membrane tilting motion is piezo-actuated, similarly to the cavity mirror mounts, and allows one to align the membrane surface parallel to the mirror reflecting surfaces or, equivalently, to align it normal to the cavity axis. Finally, the linear membrane movement along the cavity axis is also piezo actuated (using a piezo chip made by Piezomechanik, Germany) to allow remote positioning of the membrane along the cavity axis with an enhanced nanometer resolution.
The cavity-membrane mechanical assembly is visible in the photograph of Figure \ref{fig:figure3}.

\begin{figure}
	\centering
	\includegraphics[keepaspectratio=true,width=15 cm]{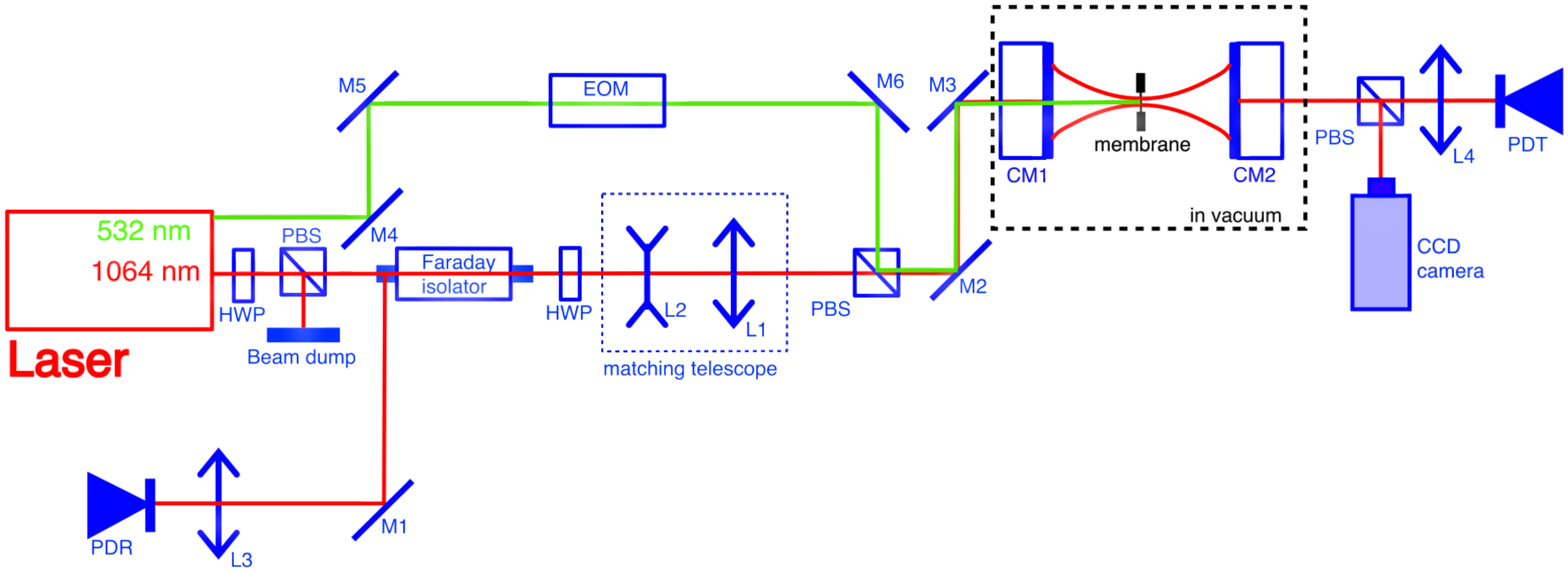}
	\caption{Schematic layout of the KWISP sensor optical system (see text for a detailed description).}
	\label{fig:figure4}
\end{figure}

The Fabry-Perot cavity is excited using a CW 1064 nm laser beam emitted by a Nd:YAG laser (Prometheus model by Innolight, Germany). This laser is also capable of emitting a second, frequency doubled, CW beam at 532 nm which is used as an auxiliary beam for alignment and for exerting an external pressure on the membrane, as described below.
Figure \ref{fig:figure4} shows a schematic layout of the KWISP sensor optical system.

\begin{figure}
	\centering
	\includegraphics[keepaspectratio=true,height=9 cm]{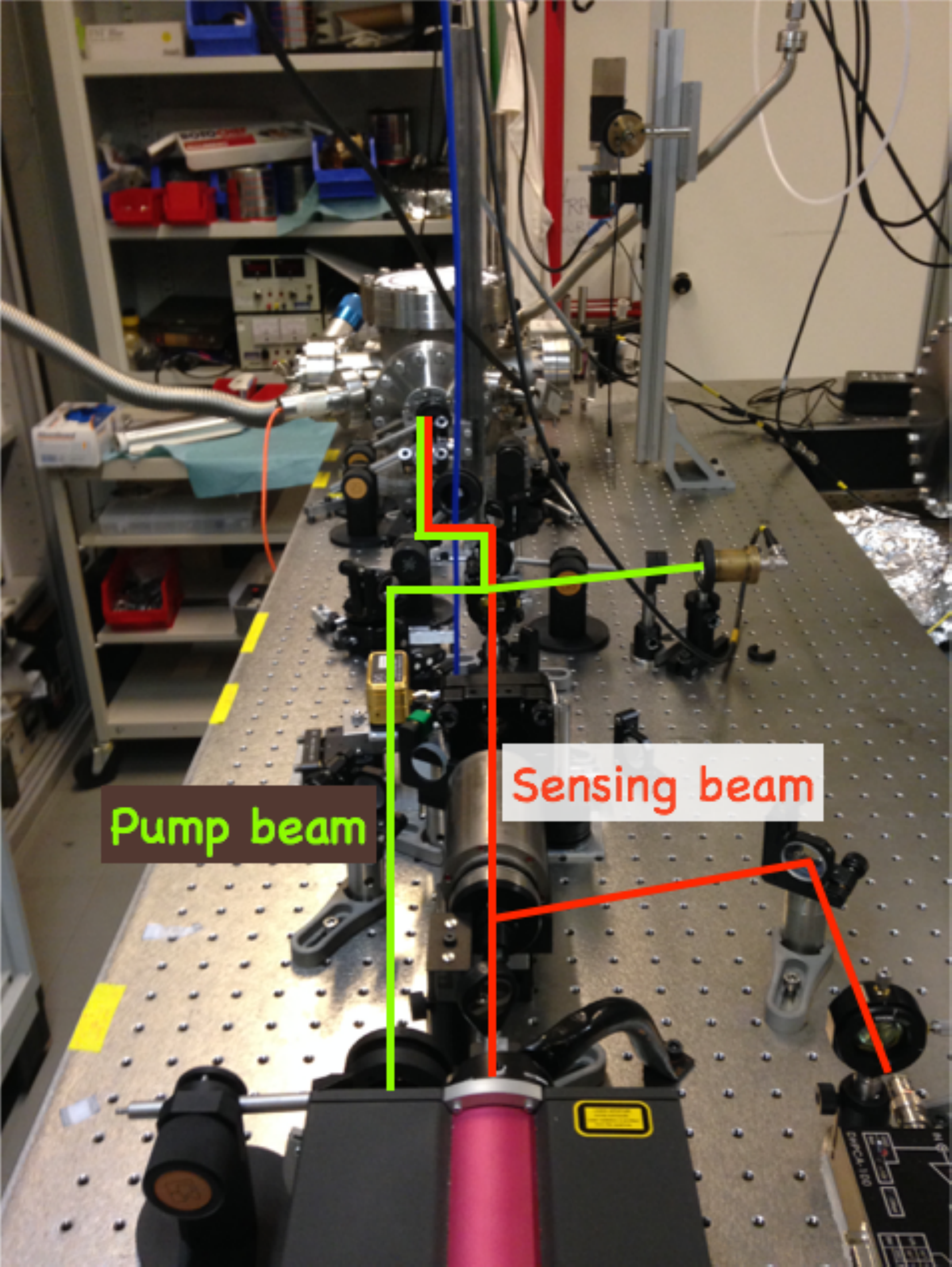}
	\caption{ Photograph of the KWISP optical bench as seen from the laser head. The vacuum chamber containing the cavity-membrane assembly is visible at back center. The paths of the two CW beams emitted by the laser are evidenced in the picture. The \emph{sensing beam}, at 1064 nm, is kept at resonance with the cavity by means of a feedback system and serves to detect membrane displacements. The \emph{pump beam}, at 532 nm, is injected in the cavity parallel to the sensing beam and exerts a pressure on the membrane by reflecting off it (see also text).}
	\label{fig:figure5}
\end{figure}

The layout shown in the figure represents a \emph{two-beam setup}: a 1064 nm \emph{sensing beam} (in red in the figure) and a 532 nm \emph{pump beam} (in green in the figure). The  \emph{sensing beam} is kept at resonance with the cavity by means of an electro-optic feedback system \cite{Cantatore1999}, and serves to detect membrane displacements. The \emph{pump beam} is injected in the cavity superimposed on the sensing beam and exerts a pressure on the membrane by reflecting off it. Immediately after leaving the laser head, the sensing beam passes through a half-wave plate (HWP) and a polarising beam-splitter (PBS), which allow a controlled attenuation of the beam intensity from a about 800 mW of CW power down to the few mW sufficient for sensor operation. The beam rejected by the PBS is directed onto a beam dump. After attenuation the beam, which is linearly polarised normal to the optical bench surface, traverses a Faraday isolator having the function of preventing the beam reflected back from the cavity input mirror (CM1) from re-entering the laser cavity and causing instabilities. A second HWP is placed after the Faraday isolator to maximise the intensity transmitted through a second PBS downstream, which allows concurrent injection of the sensing and of the pump beam into the cavity. A matching telescope, consisting of a divergent lens L2 and of a convergent lens L1, has the function of adapting the curvature of the laser beam wavefronts to match the curvature of the cavity mirrors CM1 and CM2, at their respective positions, in order to maximise the light power coupled into the cavity at resonance. The sensing beam is then injected into the Fabry-Perot cavity through a set of steering mirrors (represented by M3 and M4 in the figure). The cavity itself is formed by the two nearly-identical, multi-layer, dielectric mirrors CM1 and CM2. The membrane is also schematically represented in the figure. The measured finesse of this cavity was $\approx 30000$, both with and without the membrane.
Light exiting the cavity at resonance passes through a third PBS which further splits it into two beams: one is directed to a CCD camera, used to image the cavity spatial modes for diagnostic and alignment purposes, while the other one is focussed by the convergent lens L4 onto the surface of a photodiode (PDT). The PDT “transmission” photodiode is instrumented with a low-noise, wide-band transimpedance amplifier (Mod. DLPCA-200, by FEMTO, Germany). Light reflected from the cavity propagates backwards through the system up to the Faraday isolator, which steers it to mirror M1, through the convergent lens L3 and onto a second photodiode (PDR, equipped with a Mod. DHPCA-100 transimpedance amplifier by FEMTO, Germany). The  back-reflected light intensity is used for diagnostic purposes and also as an input for the electro-optic feedback system keeping the laser frequency continuously at resonance with the cavity. This system (not represented in Figure 4) is called the Pound-Drever-Hall feedback from its inventors, and is described in \cite{Cantatore1999} and references therein. For the purposes of the present work it sufficient to note that the system works by analyzing the back-reflected beam to obtain a signal proportional to the instantaneous difference between the laser frequency and the cavity frequency. This signal, called the \emph{error signal}, is then amplified and fed back into the laser to control its frequency. The error signal, therefore, contains the information of how the cavity frequency shifts, and it is the signal from which membrane displacement is detected. 
The 532 nm pump beam is generated inside the laser by frequency duplication of the main 1064 nm beam through a suitable non-linear crystal. This beam is amplitude-modulated by means of an electro-optic crystal excited with a sine signal at a chosen frequency (EOM, mod. 4104 by NewFocus, USA). It is then aligned on top of the main 1064 nm by means of a PBS (see Figure \ref{fig:figure4}) and, after passing through the first cavity mirror CM1, reflects off the membrane causing a time dependent force on it. With respect to the force sensor the pump beam plays, for instance, the same role that a calibration source plays for a standard radiation detector.
Figure \ref{fig:figure5} shows a photograph of the optical bench hosting the KWISP sensor as seen from the laser head.

\section{Sensor calibration and sensitivity}

\begin{figure}
	\centering
	\includegraphics[keepaspectratio=true,height=10 cm]{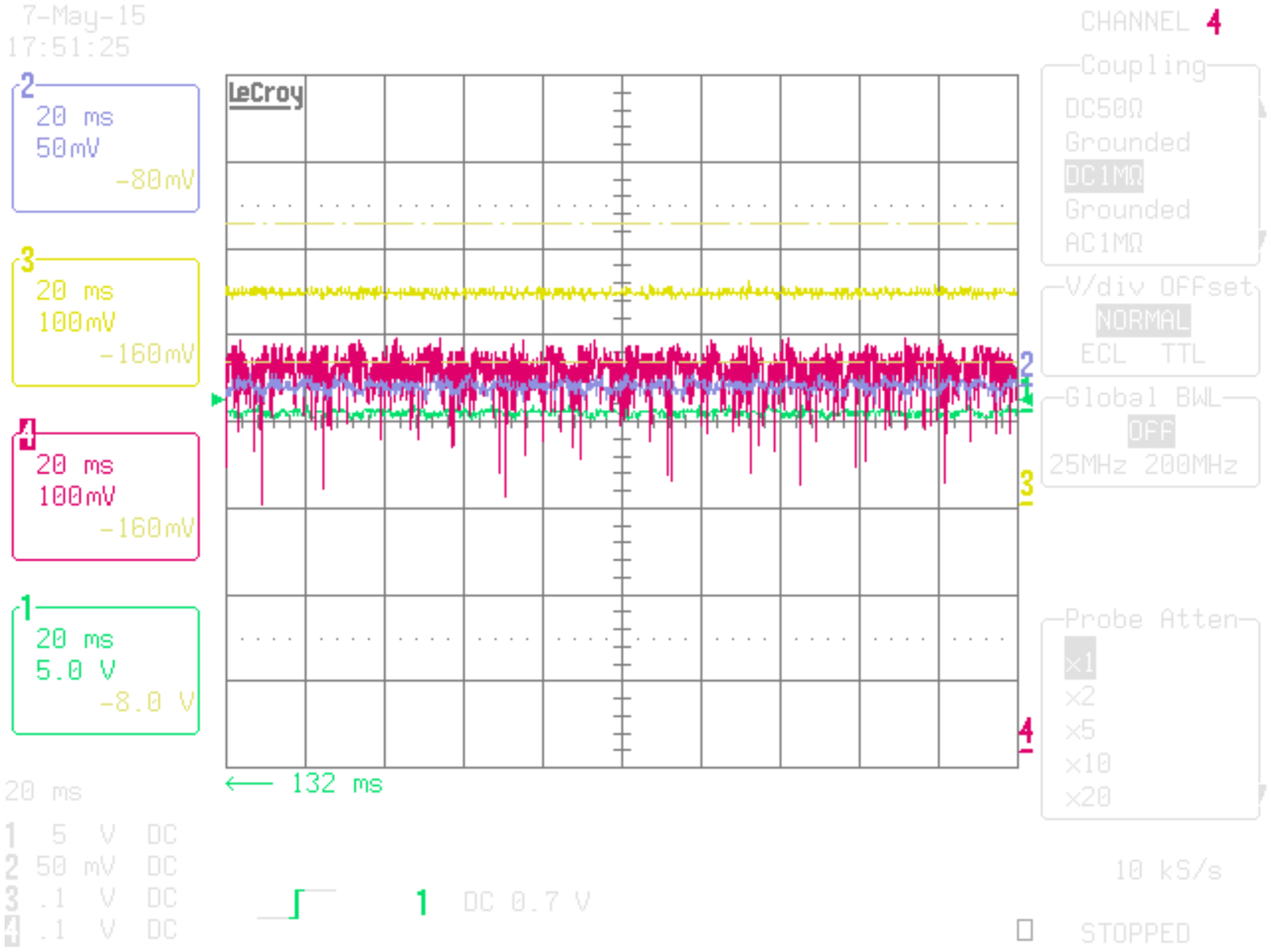}
	\caption{Sample screen shot of the oscilloscope monitoring the Fabry-Perot cavity control signals when under lock (see text). Yellow: reflected light intensity. Red: transmitted light intensity. Violet: feedback error signal. Green: feedback correction signal.}
	\label{fig:figure6}
\end{figure}

During operation the KWISP sensor is in static vacuum with a residual pressure $<10^{-3}$ mbar and the whole apparatus is at room temperature. To set the sensor in working mode the Fabry-Perot cavity is frequency locked to the laser. The locking status is monitored by observing on an oscilloscope the control signals  from the feedback circuit and the light intensities reflected and transmitted by the cavity under lock. Figure \ref{fig:figure6} shows a sample screen shot of the relevant oscilloscope traces. The top trace (yellow in the figure) gives the intensity reflected by the cavity, as recorded by photodiode PDR (see Figure \ref{fig:figure4}). When under lock, this intensity must be less than the value it has when the cavity is unlocked (yellow dashed reference line). The red trace represents the light intensity transmitted through the cavity as recorded by photodiode PDT (see Figure \ref{fig:figure4}). When under lock, this value must be greater than zero. The other two traces refer to control signals from the feedback circuit: green for the voltage being fed back to the laser to control its frequency, violet for the error signal generated by the feedback loop. Both these signals must have a small amplitude when under lock.

\begin{figure}
	\centering
	\includegraphics[keepaspectratio=true,height=10 cm]{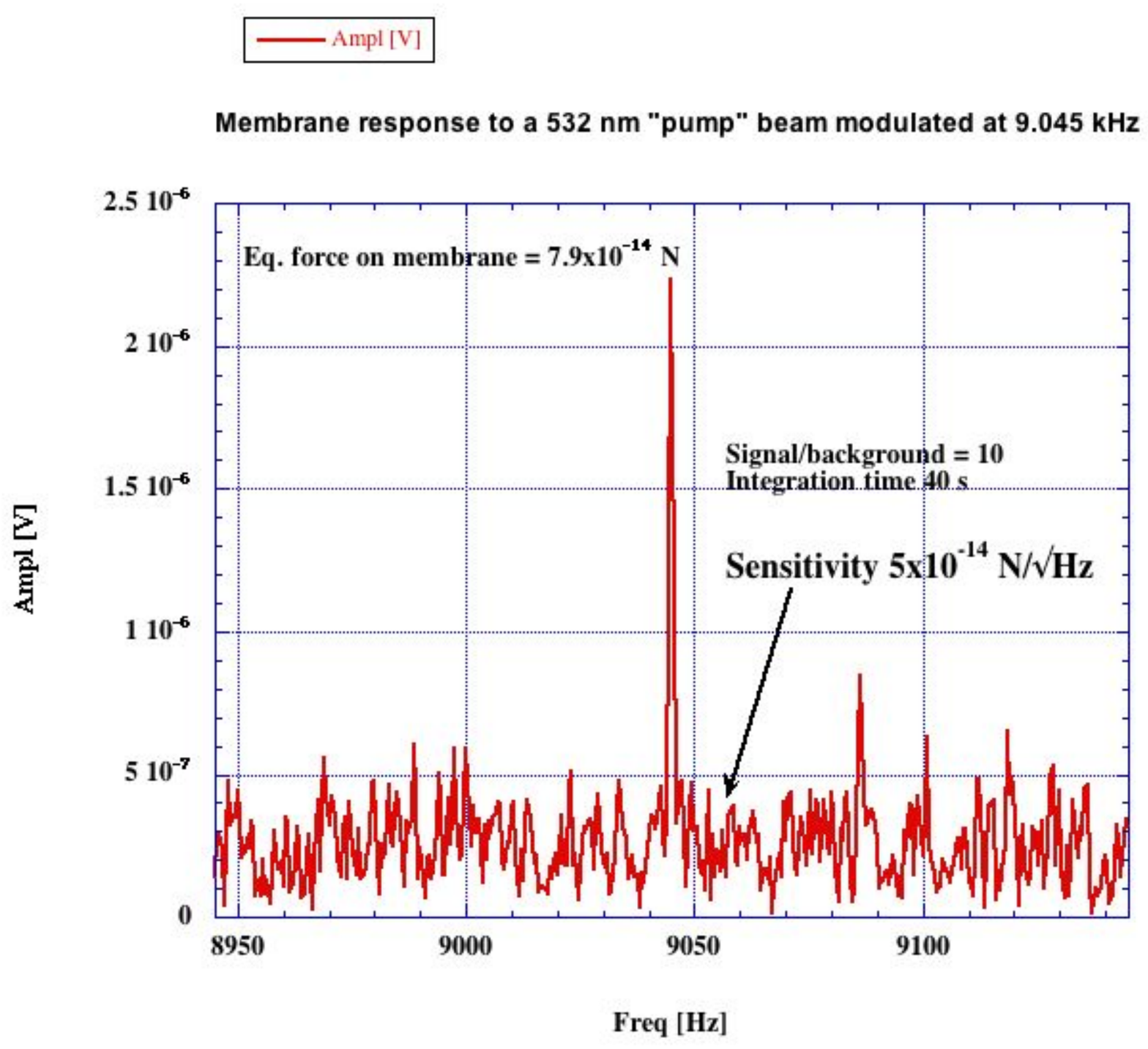}
	\caption{Power spectrum of the cavity feedback error signal when the pump beam is amplitude modulated at 9.045 kHz. The large peak indicates that the membrane is being excited by an external force, while the background is mainly due to electronic noise in the locking circuit. (see text).}
	\label{fig:figure7}
\end{figure}

The error signal is proportional to the instantaneous frequency difference between laser and cavity and its power spectrum contains the information on membrane displacements. To obtain this spectral density, the error signal is fed into an HP35660A Spectrum Analyser, which directly outputs spectral data to a data acquisition computer.
In order to achieve an absolute calibration of the sensitivity in terms of force, the pump beam is then injected onto the membrane and amplitude-modulated as described above. The pumpo light power impinging on the membrane was 600 $\mu$W, and given a measured membrane reflectivity of 0.25 at 532 nm, the light power from the pump beam reflected off the membrane was 165 $\mu$W. Taking into account a modulation index of 0.283 at 9.045 kHz, we find that the maximum amplitude of the light power reflected from the membrane in these conditions is 47 $\mu$W, corresponding to a net force of $7.9 \cdot 10 ^{-14}$ N.
The presence of this force is detected as a peak in the measured spectrum of the error signal. Figure \ref{fig:figure7} shows the power spectrum of the error signal measured when the pump beam is modulated at 9.045 kHz. The large peak visible in the plot corresponds to the amplitude of the force directly exciting the membrane, while the background is mainly due to electronic noise in the locking circuit. Since the amplitude of the exciting force is independently known, the sensitivity of the sensor can be obtained from the measured signal-to-noise ratio (SNR). With reference to Figure \ref{fig:figure7} , we find SNR = 10, therefore for the chosen integration time of 40 s, one has a force sensitivity of $5.0 \cdot 10 ^{-14}~\mbox{N}/ \sqrt{\mbox{Hz}}$. Recall that this value represents the minimum force amplitude acting on the membrane detectable in 1 s.

The membrane is actually a mechanical oscillator, and its behaviour can be modelled using a finite element analysis software to obtain its fundamental resonant frequency and its equivalent spring constant. In our present case we can assume from these simulations a spring constant of $\approx 20$ N/m (this value compares well with the 30 N/m quoted for instance in \cite{Lamoreaux2008}). Then the equivalent displacement sensitivity of the KWISP sensor is $2.5 \cdot 10 ^{-15}~\mbox{m}/ \sqrt{\mbox{Hz}}$ at 9.045 kHz.

\begin{figure}
	\centering
	\includegraphics[keepaspectratio=true,height=9 cm]{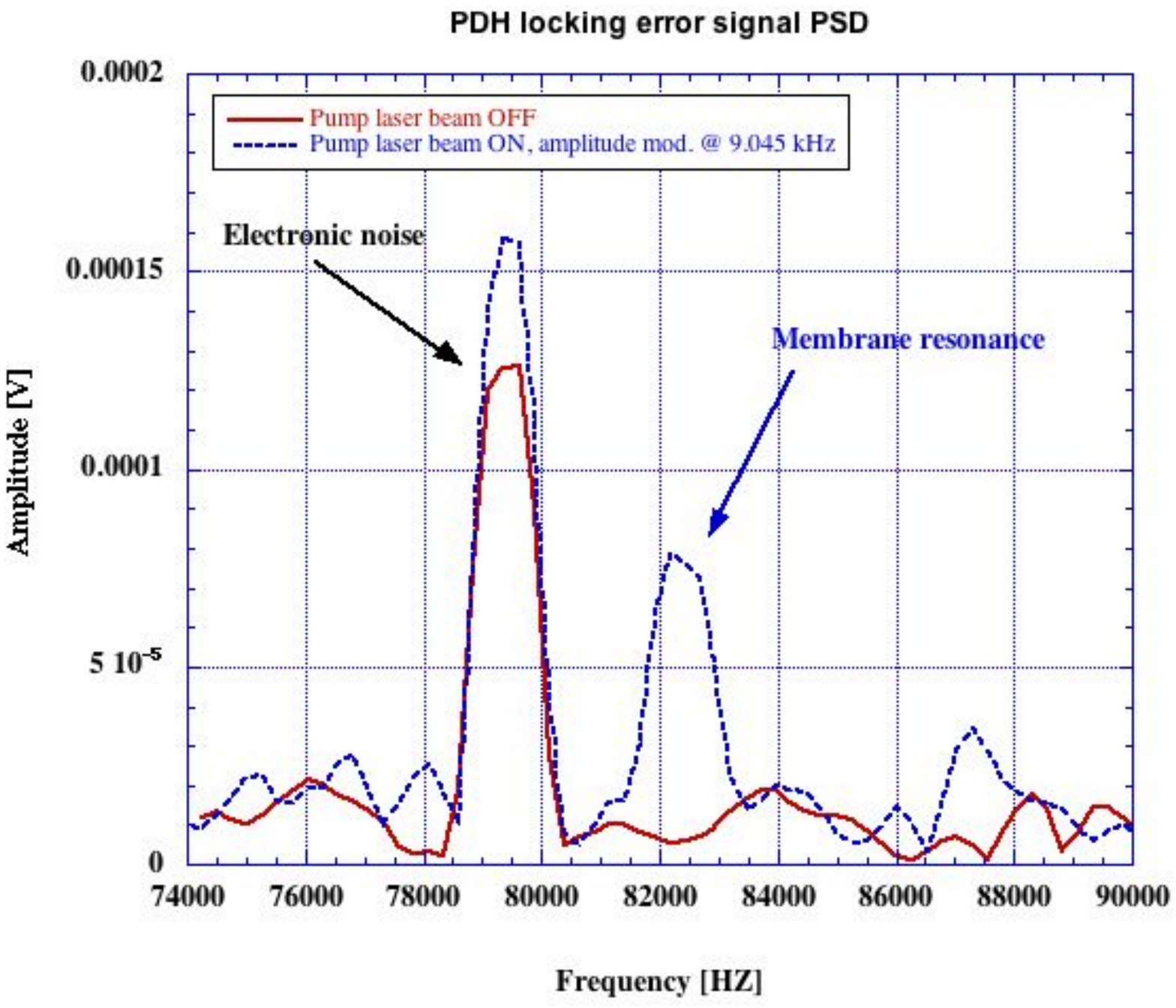}
	\caption{Plot of the power spectrum of the error signal around 82 kHz with the pump beam OFF (red curve) and ON (blue curve). Recall that the pump beam is in this case amplitude-modulated at 9.045 kHz (see also text).}
	\label{fig:figure8}
\end{figure}

The mechanical oscillator behaviour of the membrane and its fundamental resonant frequency can be directly measured with the pump beam technique.
Figure \ref{fig:figure8} shows the power spectrum of the cavity feedback error signal around 82 kHz when the membrane is excited by a pump beam amplitude-modulated at 9.045 kHz, that is, off the expected fundamental mechanical resonance frequency of the membrane. The red curve in the plot of Figure \ref{fig:figure8} gives the power spectrum measured when no pump beam is present, showing the presence of a spurious peak generated in the electronics. The blue curve represents the spectrum measured with the pump beam on, and shows the appearance of a peak at $\approx 82.5$ kHz. This peak is due to energy from the pump beam coupling to the fundamental mechanical resonant mode of the membrane.

Given the measured fundamental mechanical frequency of the membrane, one can further investigate the behaviour of the membrane by exciting it with a pump beam amplitude-modulated at frequencies around the mechanical resonant frequency. Preliminary measurements indicate a membrane quality factor of $\mbox{Q}_{meas}\approx3000$ and a sensitivity of $1.5 \cdot 10 ^{-14}~\mbox{N}/\sqrt{\mbox{Hz}}$. The measured Q factor is lower than quality factors in excess of $10^{5}$ routinely found in the literature \cite{Zwickl2008,Joeckel2011}. We attribute our poorer quality factor from these preliminary measurements to an insufficiently low residual gas pressure in the sensor vacuum chamber, which dampens membrane oscillations. The sensitivity, on the other hand, is practically at the 300 K thermal limit estimated using the measured Q \cite{Lamoreaux2008}.

\section{Conclusions}

We have built an ultra-sensitivite, opto-mechanical force sensor, called KWISP, which is now in operation in the optics laboratory at INFN Trieste. The sensor is based on a thin micro-membrane inserted in a Fabry-Perot optical resonant cavity. An absolute calibration of the force sensitivity of this device has been obtained by exciting the membrane with an amplitude modulated light beam. The off-resonance measured force sensitivity is $5.0 \cdot 10 ^{-14}~\mbox{N}/ \sqrt{\mbox{Hz}}$, corresponding to a sensitivity to membrane displacements of $2.5 \cdot 10 ^{-15}~\mbox{m}/ \sqrt{\mbox{Hz}}$. Note that this distance is comparable to the average radius on an atomic nucleus. Preliminary measurements around resonance indicate a thermally-limited force sensitivity of $1.5 \cdot 10 ^{-14}~\mbox{N}/ \sqrt{\mbox{Hz}}$, corresponding to a displacement sensitivity of $7.5 \cdot 10 ^{-16}~\mbox{m}/ \sqrt{\mbox{Hz}}$. This figure is comparable to the displacement sensitivities achieved by large interferometric gravitational wave detectors \cite{Virgo2014}, while our sensor is of course not as sensitive in terms of gravitational waves, given the much shorter length of the Fabry-Perot cavity.
For the KWISP sensor under better vacuum it is reasonable to  expect Q $\approx 10^{5}$ \cite{Lamoreaux2008}, giving a thermally-limited sensitivity of $\approx2.5 \cdot 10 ^{-15}~\mbox{N}/ \sqrt{\mbox{Hz}}$. A further factor of >100 in sensitivity could be gained by cooling the membrane from room temperature down to sub-K temperatures (30 mK for instance). In this case the projected sensitivity is as low as $\approx8.0 \cdot 10 ^{-18}~\mbox{N}/ \sqrt{\mbox{Hz}}$.
An immediate application we foresee for the KWISP force-sensor is in the search for Chameleon-type scalar WISPs \cite{Baum2014} to be conducted shortly at CAST \cite{Cantatore2015}. There, in CAST, one will exploit both the sun-tracking capability of the moveable magnet carriage, and the presence of an X-ray telescope, which acts also as a focussing device for Chameleons.
The extreme sensitivity of the KWISP sensor to tiny displacement makes it also suitable and very attractive for applications in the field of the experimental study of interactions at short distances, with immediate impact on the physics of extra-dimensions and quantum gravity \cite{Geraci2008,Geraci2010,Arvanitaki2015,Baharami2015}.

\paragraph{Acknowledgements}
We thank D. Vitali, R. Natali and G. Di Giuseppe from the University of Camerino, Italy, for their support and their useful suggestions. We also thank A. Arvanitaki and S. Dimopoulos for the fruitful discussions. This work was partially supported by grants no. 13.12.2.2.09, University of Rijeka, and no. 533-19-14-0002, Croatian Ministry of Science, Education and Sports.

\end{document}